# Toward a New Policy for Scientific and Technical Communication: the Case of Kyrgyz Republic

Nurlan Djenchuraev, IPF policy fellow, Open Society Institute (Djenchuraev@policy.hu)

## Table of Contents

## Abstract

The objective of this policy paper to formulate a new policy in the field of scientific and technical information (STI) in Kyrgyz Republic in the light of emergence and rapid development of electronic scientific communication. The major problem with communication in science in the Republic is lack of adequate access to information by scientists. An equally serious problem is poor visibility of research conducted in Kyrgyzstan and, as consequence, negligible research impact on academic society globally. The paper proposes an integrated approach to formulation of a new STI policy based on a number of policy components: telecommunication networks, computerization, STI systems, legislation & standards, and education & trainings. Two alternatives were considered: electronic vs. paper-based scientific communication and development of the national STI system vs. cross-national virtual collaboration. The study results in suggesting a number of policy recommendations for identified stakeholders.

## Background

Revolutionary development of the information and communication technologies sector in recent years resulted in such a great leap forward in scientific communication that is considered to be comparable by significance with Gutenberg's invention of the movable type printing press in 15[th] century. The arrival of new information technologies affords breathtaking opportunities with respect to rapidity of scientific information exchange, information retrieval, and multimedia presentation, but also brings up serious issues concerning data archiving, intellectual property aspects, and peer review.

The development and access to electronic scholarly networks is distributed extremely unevenly among and within nations. While in industrialized countries widespread Internet access to libraries, databases, electronic academic discussion lists, and electronic communication are ordinary, in developing ones the situation is quite the contrary.

The question whether the scientific information gap between Kyrgyzstan and developed countries will be narrowed or widened with the emergence of Internet technologies heavily depends on formulation and implementation in Kyrgyzstan of integrated scientific and technical information policy.

## Statement of Purpose

The major goal of this paper is to formulate a new scientific and technical information (STI) policy in Kyrgyz Republic in the light of emergence and rapid development of electronic scientific communication and provide recommendations to various stakeholders involved.

The intended audience for this study includes the policymakers, international donor agencies, academic society, librarians, science publishers and all those interested in STI policy issues in Kyrgyz Republic.

It is hoped that this paper will serve as a basis for adoption of a new national policy in the field of STI system in the country.

## Scope

Although the scope of this study is scientific and technical communication in Kyrgyzstan, issues associated with global trends in scientific communication will also be discussed. An emphasis will be mostly placed on electronic communications, and transfer from paper to Internet based scientific communications. Two basic issues under discussion will be how to improve access to scientific information to scientists in Kyrgyzstan and how to provide global visibility of research made by Kyrgyz scientists.

Due to close historical roots with other former Soviet Union countries, considerations and recommendations represented in this document may be to some extent brought into correlation to scientific communication in other former Soviet Union countries.

## Methodology

The methodology of this study included analysis of the global tendencies in scientific communication on the basis of facts obtained from literature references. Trends in scientific and technical communication in Kyrgyzstan were traced as a result of both full-bodied study of literature and interviews, quantitative survey and focus groups with government officials, scientists, publishers and librarians. More than 100 scientists in Bishkek, Osh and Kara-Kol were surveyed. More detailed information about methodology and results obtained can be found in the background study (Djenchuraev 2004).

## The Problem

As in many former Soviet Union countries, after gaining independence in 1991 science in Kyrgyzstan experienced a severe crisis. State Program on Reforming of Science in Kyrgyz Republic for 2003-2005 provides some figures on deteriorating status of science in the Republic: the share of expenses for research and development in the gross domestic product was gradually decreasing: in 2001 it amounted only 0.1%; number of people involved in science has dropped to about 4,600 people. Declining prestige of science, lack of funds for scientific research and decreasing wages of scientists, ageing academic community, deteriorating scientific communication – this is only a partial list of challenges facing science in Kyrgyzstan.

The scientific communication problem is one of the most acute problems as was found as a result of several focus groups with scientists and a quantitative survey of academic community. (Djenchuraev 2004).

As could be expected, scientific communication in the country is heavily dependent on paper-based technologies and lacking a book or journal in the library is a most considerable barrier for about 62% scientists as shown in Fig.1.

A less predictable issue turned out to be a language (English) barrier. About 42% of scientist surveyed named it as an obstacle; it is worth noting that it is much lower for younger scientists (20-29 y.o) – 28% as compared with other age groups – from 44% to 55%.

The high price of accessing data is the third most important obstacle to scientific communication in Kyrgyzstan (30%).

Surprisingly, barriers related to computer and Internet turned out to be not so high: only 22% mention poor knowledge of computer and 24% have problems with searching information in the Internet.

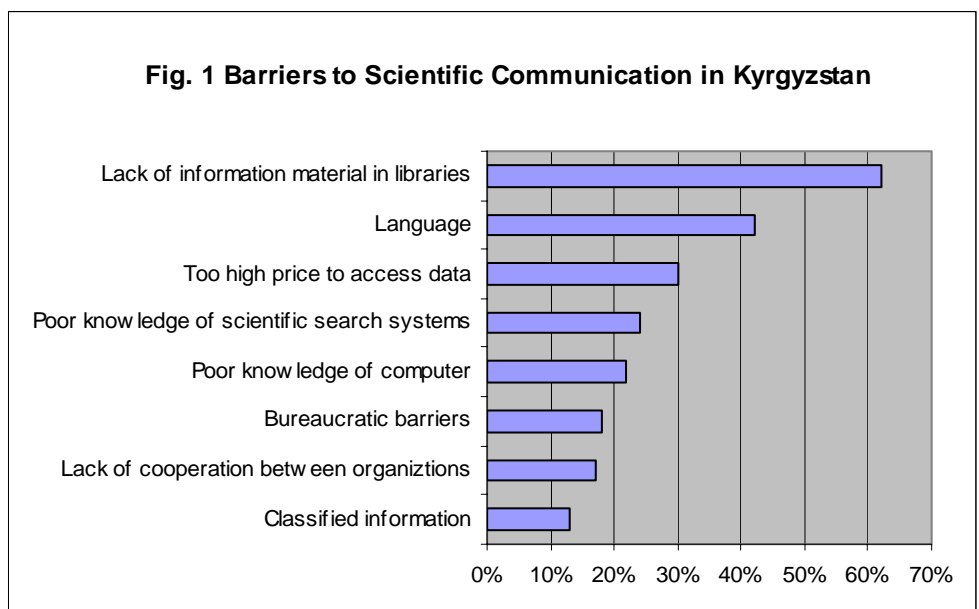

Source: (Djenchuraev 2004)

As for the other obstacles such as classified information or bureaucratic barriers– they appear not so important.

The crisis in science affected all actors involved in scientific communication, namely scientists (as readers and writers), publishers, libraries turning them into marginalized groups:

- Scientists as writers experience difficulties in having adequate research impact or "visibility" of their publications and scientific communication; as a result Kyrgyz science is a sort of "lost science" for global academic community. Scientists as readers do not have

- adequate access to local or worldwide science literature due to low circulation of the former one and lack of the latter one.
- Science publishers are practically not involved in scientific publishing.
- Libraries experience considerable difficulties in subscription to science magazines and books due to lack of proper funding.

It is worth noting that different lines of attack on the problem were outlined in policy documents: National Strategy "Information and Communication Technology for the Development of Kyrgyz Republic and National Poverty Reduction Strategy 2003-2005 within "The Comprehensive Development Framework of the Kyrgyz Republic to 2010", the Concept on Reforming of Science in Kyrgyz Republic for 1999-2005, the State Program on Reforming of Science in Kyrgyz Republic for 2003-2005, and the Agreement on Free Access and Exchange Procedure for Open Scientific and Technical Information signed by CIS countries on September 11, 1998. Legislative framework includes the Law on Science and Basics of Scientific and Technical Policy (1994) and Law on Scientific and Technical Information System (1999). A standard on electronic editions which has been laid down by several CIS countries came into force July 1, 2002. Moreover, a number of international donor agencies and NGOs implement various projects aimed at enhancing scientific communication in Kyrgyzstan, for example eIFL.net assists libraries in achieving access to scientific electronic resources, IREX assists scientists in their access to Internet and provides basic computer trainings, INTAS funds e-library action with the aim of providing access of researchers to electronic resources, Virtual Silk Highway project funded by NATO aimed at providing global Internet connectivity to science and education community in Kyrgyzstan.

There is no question that the above policies and initiatives may have an important impact on scientific communication in Kyrgyzstan. The results of implementation of many programs are tangible and beneficial. Nevertheless, many more advances could be achieved if there would be a clear notion of the current status of scientific communication in Kyrgyzstan, who are the policy stakeholders, what are the major obstacles to communication of scientists, what is an "ideal picture" of STI system in the country, and so on. Moreover, there is no integrated national scientific information policy taking into account rapid penetration of Internet technologies in the field of scientific communications and globalization of science.

# Policy Alternatives for Scientific Information

## *Electronic vs. Print Scientific Communication*

As early as in the beginning of 90-s the STI system in Kyrgyzstan was a part of the State System of Scientific and Technical Information (GSNTI) of USSR. GSNTI was based on the principals of the leading role of the state in scientific information management and its centralized processing. On the whole, the system was considerably based on paper technologies. Due to awkwardness and rigid structure GSNTI was not able to begin timely change-over to microcomputer base and due to lack of funding it failed in the beginning of 90-s. In post-Soviet times many information institutions and research libraries were deteriorating being on the verge of survival. The amount of subscription for books and scientific journals has dropped by many times. In essence, it was similar to "serials crises"[1] in the West, but in much larger-scale.

All these events took place against the backgrounds of rapid development of electronic communications in the West and their penetration in the field of science and education. Transfer to network technologies, emergence of new models of scientific communications such as e-prints, e-journals, videoconferencing, listservs and forums, etc and attempts to provide free access to scientific publications have been occurring for very short period of time.

---

[1] A situation when libraries can ill afford to maintain the collection of journals to which they subscribe due to high prices

The end of 90-s was marked by increasing penetration of ICT technologies in scientific communications in Kyrgyzstan as well. The paucity of the data on scientific communication in Kyrgyzstan encouraged making a study of the current state of art in this field. (Djenchuraev 2004). Some findings of this study are demonstrated below:

- Although traditional informal scientific communications play a significant role at local level – 52% of scientists surveyed used this channel of information over the past year - it is not so evident for communications at longer distances where electronic technologies offer several advantages over expensive traditional contacts at conferences, seminars, etc. The share of scientists using e-mail for foreign scientific communication is equal to communicating directly. This is due to high level of e-mail availability – presently 76% of scientists have access to private or organization e-mail; about one third of all scientists belongs to advanced users (frequency of e-mail usage 1 time per day or more), and about two thirds – to a group with average e-mail usage (from 2-3 times per week to 1 time per week). It is anticipated that the next stage of the development of electronic informal scientific communications will be transformation from traditional "invisible colleges" to cyberspace colleges; currently only 10-12% of scientists are subscribed for newsgroups and forums.
- Despite the fact that formal communications are mostly paper-based (72% of all scientists currently use traditional libraries), electronic communications are gaining wider recognition, especially for scientific e-journals (38%). It comes as no surprise as there are several thousands on-line electronic scientific journals. Nonetheless, it is necessary to note that publication in electronic journals is much less as compared with publication in traditional journals resulting in "invisibility" of local research.

With all these issues in mind the next question to consider will be - what should be the policy option to improve current scientific communications in Kyrgyzstan: traditional paper-based or electronic communications? The answer is not so evident as may appear at first glance and requires deeper analysis.

It is well known that there are three forms of scientific communication: oral (spoken), documentary (print, written) and electronic. While oral communication is based on verbal and nonverbal natural communication channels which can not be replaced, documentary and electronic communication compete with each other. If communication barriers for electronic communication are lower than documentary communication, the latter will be squeezed out to the periphery of social communications (Sokolov 2002).

Nevertheless, the posture of documentary communication is still strong due to enormous inertia of the scientific publishing system (Odlyzko 2002).

Advantages and disadvantages of paper and electronic communications policy options are demonstrated in Table 1.

Table 1. Analysis of policy options: paper-based vs. electronic communications

|  | Advantage | Disadvantage |
|---|---|---|
| Paper-based communication | Experience in organization, no need in expensive infrastructure | Inefficiency, tardiness, overmanning of operating personnel, high cost of subscriptions |
| Electronic communication | High speed, efficiency, flexibility, multimedia presentation, very low cost | Considerable expenses to telecommunication infrastructure, computerization, development of local electronic databases and access to global ones, digitization. Need to reconsider legislative framework and standards. Training of scientists, librarians, and science publishers. |

Summing up, it can be said that electronic scientific communications show considerable promise but balanced policy is needed not to destroy existing paper scientific communication in a revolutionary way. As things now stand a combination of policy options can be proposed for the development of scientific communication in Kyrgyzstan, although development of electronic scientific communications should definitely prevail.

### *National STI System vs. Cross-National Virtual Collaboration*

Analysis of global tendencies in scientific communication enabled to identify two potential directions for the development of scientific and technical communication in Kyrgyzstan:

1. **Developing national scientific and technical information system (STI policy)**. This policy points towards setting up national scientific information system including academic, university and sectoral science networks and their linking with global scientific information system. It focuses mostly on the uniform development of the national STI system on the whole.
2. **Developing virtual laboratories or collaboratories (VL policy).** This policy is a reflection of a global trend to globalization of science. It is based on virtual collaboration of scientific teams and laboratories in global research networks through geographically distributed scientific research across the digital divide. The focus of this policy is chiefly on the development of cross-national "virtual" collaboration in leading-edge science. Unlike traditional international collaboration virtual laboratories use information channels of such a bandwidth to enable geographically distributed teams to conduct full-scale research as if they work in one laboratory.

Table 2 gives a comparative analysis of policy options regarding organization of scientific communication.

Table 2. Analysis of policy options: STI vs. VL

|  | Advantage | Disadvantage |
|---|---|---|
| STI policy | Balanced development of national science resources in general, benefits for more scientists. | High probability of "lost science", no breakthrough research centers. |
| VL policy | Opportunity to participate in global innovative research, no brain-drain, international prestige, education of skilled young scientists. | Considerable expenses to telecommunication infrastructure and computerization, lack of experience. |

There is an experience of successful collaboratory in a developing country. Virtual institute of 35 laboratories in Brazil participated in global Human Genome Project competing with developed countries. The result of this project was increasing prestige of Brazil science, new investments in scientific research and training of more than 200 young geneticists (Collins 2000).

# Components of a New Scientific and Technical Information Policy

Recent OECD study (OECD 1998) proposes three different roles the government can play in the ICT science policy:
    1) Support of technical infrastructure used by scientists;
    2) Development of a legislative framework which regulates intellectual property rights, access to information, etc;
    3) Repletion of ICT needs such as ICT training of scientists, establishment of electronic databases.

Developing this model and taking into account a global trend to transferring scientific communication from paper to electronic media the new STI policy can be formulated (Djenchuraev 2004). The components of this policy embrace the following:

- **Telecommunication infrastructure.** ICT infrastructure is a core of electronic scientific and technical communication system. Technical capabilities of telecommunication networks define in many ways the speed and reliability of access to scientific resources and level of communication among scientists. Public policy in telecommunication field is a decisive component of information support of science as stated in National Poverty Reduction Strategy 2003-2005.
- **Computerization**. Implementation of a policy to increase computerization of research organizations, libraries, publishers. The state should stimulate manufacturing, import and usage of modern computer systems and software.
- **STI system**. The policy in the field of STI systems should be corrected in such a way to the full extent use potential of ICT technologies including both formal and informal electronic communications. Although electronic communications should play a dominant role in the new STI policy, documentary communications should be maintained at minimum necessary level as well. Both national system of scientific and technical information and establishing geographically distributed research should be developed.
- **Human resource development**. This includes education policy to minimize as far as possible scientific communication barriers caused by human factor. The policy is aimed at training ICT specialists as well as scientists, librarians and science publishers in the field of electronic STI systems.
- **Legal framework development**. This component represents improvement of legislative framework and development of standards in the field of telecommunications, information science and STI systems. Standards and security of scientific and technical information is also involved in this component.

## *Policy Stakeholders*

For the purposes of this paper stakeholders can be defined as "any group or individual who have a stake in a scientific (and technical) communication policy because they affect or are affected by government decisions".

Several groups of stakeholders can be distinguished at national and international levels as shown in Table 3.

Table 3. Groups of national and international stakeholders in scientific communication system in Kyrgyzstan

| National stakeholders | International stakeholders |
|---|---|
| Group 1. Government<br>State Agency on Science and Intellectual Property<br>National Academy of Sciences of Kyrgyz Republic<br>Ministry of Transport and Communications<br>Ministry of Education and Culture<br>Council on ICT under the President of Kyrgyz Republic<br>Ministry of Justice<br>Ministry of National Security | Group 1. Governments of other countries (CIS, et al ) |
| Group 2. Academic Community<br>Individual scientists, post-graduate students, associations of scientists. | Group 2. International Scientists<br>Individual scientists, post-graduate students, associations of scientists |
| Group 3. Research Libraries, University Libraries | Group 3. International research and university libraries |
| Group 4. Science Publishers<br>Science publishing houses, editors, referees | Group 4. International science publishers<br>Group 5. International Donor Organizations & Foundations<br>OSI, ICSU, IREX, UNESCO, OCED, UNDP, etc |

Involvement of the identified stakeholders in implementation of policy recommendations is shown later.

# Conclusion and Recommendations

The system of scientific and technical communication in Kyrgyz Republic is currently in transformation from paper to electronic media. Scientific communication based on ICT technologies could potentially provide the best solution to improving communication; however, there are a number of obstacles: telecommunication infrastructure is underdeveloped, the level of computerization of academic society, libraries and publishers is low, there is lack of adequate electronic scientific resources, correspondent legislation and standards as well as training and educational programs.

A new scientific and technical information policy based on Internet technologies should take into account proportional development of all correspondent policy components, namely:
- Telecommunication infrastructure;
- Computerization;
- Scientific and technical communication systems;
- Legislation;
- Education & training.

International Council for Science (ICSU) suggests the following list of actions required to protect public domain information (ICSU 2003):
1) All governments should adopt policies to ensure that data produced from publicly funded research remains openly available to the largest extent possible;
2) Involve scientists as major stakeholders in the information society, in the discussion and development intellectual property rights and copyright legislation – at both national and international level;
3) Appropriate national policies should be developed to facilitate the exchange of scientific information;
4) Develop special programs for scientific collaboration across the digital divide, thereby facilitating exchange of scientific information and knowledge;
5) Promote different models of scientific information production and dissemination, including those particularly appropriate for developing countries.

Taking into account the above list of action the policy recommendations for this study are summarized in Table 4 as four sets of issues: development of the STI framework, improving availability of scientific information to scientists in Kyrgyzstan, increasing visibility of research conducted in Kyrgyzstan and development of global virtual collaboration:

Table 4. Policy recommendations

| Recommendations | Stakeholder |
|---|---|
| **Development of the STI framework** ||
| **STI Policy:** It is recommended that Government (and other stakeholders) develop a detailed program of transfer to electronic scientific communications taking into account proportional development of all policy components: telecommunication policy, computerization policy, STI systems policy, legislation and education. Due balance should be kept between electronic and paper communications. | Gov (NAIP, MTC, MEC, MNC, AdP, NAS), libraries, |
| **Telecommunication networks:** It is recommended that Government should proceed with telecommunication development program. Specific telecommunication needs of STI system should be discussed by NAIP and MTC. | Gov (NAIP, MTC) |
| **Computerization:** It is recommended to introduce measures to contribute to faster computerization of research institutes, universities and libraries. This component should be | Gov (NAIP, MTC), Don |

|   | | |
|---|---|---|
| | harmonized with telecommunication networks development. | |
| | **Development of STI systems.** It is advisable to develop an updated structure of STI system in Kyrgyzstan based on dual paper/electronic system and develop a strategy of increasing of electronic communication component. | Gov |
| | **Development of legislation framework and standards.** It is advisable for Government to conduct a study of current legislation and standards in the field of STI. A conference on legislation and standards involving local and international stakeholders may be organized by NAIP to discuss intellectual property rights, security of information, etc. especially related to the electronic environment. | Gov(NAIP, MTC, MNS, MJ, NAS), Uni, Lib, Don, JK |
| | **Development of human potential.** It is recommended that Government should stimulate universities and other organizations to establish education programs for specialists developing and maintaining STI systems. Equally, it is encouraged to establish trainings for scientists, librarians and science publishers. International donor organizations have been proposed to assist in the development and organization of training programs such as academic English for scientists, electronic publishing, etc. | Gov (NAIS, MEC), Don |
| colspan=2 | **Improving availability of scientific information to scientists in Kyrgyzstan** | |
| | Support of access to global databases (for example, EBSCO) considering priority science fields. It is recommended that Government (NAIP) allocated financial resources for subscription to EBSCO for next years. It is recommended for international aid agencies to provide open access to other full-text databases. | Gov (NAIP), Don |
| | Information database should be composed including all open resources of scientific information available for scientists in Kyrgyzstan and published at NAIP web-site. | Gov (NAIP), Lib |
| | A subscription to databases including Russian resources is strongly recommended. | Gov (NAIP), Don |
| | Trainings for scientists, librarians, science publishers to remove language and computer-related barriers. | Gov (NAIP), Uni, Don |
| colspan=2 | **Increasing visibility of research conducted in Kyrgyzstan** | |
| | It is recommended to develop and enforce enactments providing obligatory publication in Internet of data obtained as a result of research work funded by tax-payers in Kyrgyzstan. This provision should also involve all research conducted within development projects funded through international credits and should be stipulated in loan documents. It is recommended to negotiate with international donor organizations concerning potential transfer of their final reports for completed earlier projects. In case of research carried out on a grant basis, the decision on open publication is made by a donor organization; however it should be informed about this possibility. | Gov (NAIP, MJ), Don |
| | It is recommended for NAIP to establish a Center of Electronic Reports and publish there all the publicly funded research in the open access. | Gov (NAIP) |
| | Retrospective and prospective digitization of print scientific materials and publishing them for open access where possible: special programs should be developed to transfer to digital form scientific journals, books, dissertations, and reports beginning from those which are under threat of destruction due to storage conditions or rare editions. intellectual property and standardization issues should be paid special attention at that. Setting up institutional e-print archives. | Lib, Gov, Pub, Sci, Don, |
| | It is recommended that the Government should foster universities, NAS and publishing houses to establish scientific electronic journals or publish electronic versions of print journals. Government funding should be provided for establishment of e-journals. A model open access scientific e-journal could be "Nauka I novye technologii" published by NAIP. International aid agencies can assist in setting up such journals through funding. It is advisable for universities or National Academy of Sciences to setup an electronic scientific journal in English to publish translations of the best papers printed in local journals. | Gov (NAIP, NAS), Uni, Pub, Lib, Don |
| | It is advisable that National Attestation Commission would acknowledge publication in approved peer-reviewed scientific electronic journals as equal to print ones for graduate students who plan to defend their dissertations. | Gov (NAK) |
| | It is recommended that international donor agencies would organize trainings for scientists on alternative forms of scientific publishing. | Don |
| colspan=2 | **Development of global virtual collaboration** | |
| | It is recommended for the Government to promote international virtual collaboration across the digital divide by supporting, on inter-governmental level, scientist's initiatives on establishment of international collaboratories including scientific teams from Kyrgyzstan. It is therefore best for government and donors to provide permanent funding for the establishment and development of a model collaboratory in one of the priority science fields. | Gov (NAIP, NAS), ISci, Don |

Stakeholders: Don – International donors; ISci – scientists abroad; JK – Jogorku Kenesh (Parliament); NAS – National Academy of Sciences; NAIP – National Agency on Intellectual Property; MJ – Ministry of Justice; MTC – Ministry of Transport & Communications; MEC – Ministry of Education & Culture; MNS – Ministry of National Security; NAK – National Committee on Attestation; Pub – science publishers; Sci – scientists; Uni - universities

# Abbreviations

| | |
|---|---|
| CIS | Commonwealth of Independent States |
| GS NTI | State system of scientific and technical information; |
| ICT | Information and communication technologies; |
| STI | Scientific and technical information; |
| VL | Virtual laboratories (collaboratories). |